\documentclass[aps,prd,twocolumn,showpacs,psfig,superscriptaddress,longbibliography]{revtex4-2}
\usepackage{amsfonts}
\usepackage{mathrsfs}
\usepackage{amsmath}
\usepackage{color}
\usepackage{natbib}
\usepackage{textcomp}
\usepackage{graphicx}
\usepackage{bm}
\usepackage{amssymb}
\usepackage{xspace}
\usepackage{epstopdf}
\usepackage{dcolumn}
\usepackage{longtable}
\usepackage{multirow}
\usepackage[colorlinks=true, letterpaper=true, pdfstartview=FitV, linkcolor=blue, citecolor=blue, urlcolor=blue]{hyperref}
\usepackage{float}

\makeatletter

\newcommand{\Rmnum}[1]{\expandafter\@slowromancap\romannumeral #1@}
\makeatother

\begin{document}

\title{$P$-orbital magnetic topological states on square lattice}

\author{Jing-Yang You}
\affiliation{Kavli Institute for Theoretical Sciences, and CAS Center for Excellence in Topological Quantum Computation, University of Chinese Academy of Sciences, Beijing 100190, China}
\affiliation{Department of Physics, Faculty of Science, National University of Singapore, 117551, Singapore}

\author{Bo Gu\textsuperscript{a}\thanks{\noindent\textsuperscript{a}Address of}}
 \email{gubo@ucas.ac.cn}
 \affiliation{Kavli Institute for Theoretical Sciences, and CAS Center for Excellence in Topological Quantum Computation, University of Chinese Academy of Sciences, Beijing 100190, China}
\affiliation{Physical Science Laboratory, Huairou National Comprehensive Science Center, Beijing 101400, China}

\author{Gang Su}
\email{gsu@ucas.ac.cn}
\affiliation{Kavli Institute for Theoretical Sciences, and CAS Center for Excellence in Topological Quantum Computation, University of Chinese Academy of Sciences, Beijing 100190, China}
\affiliation{Physical Science Laboratory, Huairou National Comprehensive Science Center, Beijing 101400, China}
\affiliation{School of Physical Sciences, University of Chinese Academy of Sciences, Beijing 100049, China}

\begin{abstract}
Honeycomb or triangular lattices were extensively studied and thought to be proper platforms for realizing quantum anomalous Hall effect (QAHE), where magnetism is usually caused by $d$ orbitals of transition metals. Here we propose that square lattice can host three magnetic topological states, including the fully spin polarized nodal loop semimetal, QAHE and topologically trivial ferromagnetic semiconductor, in terms of the symmetry and $k\cdot p$ model analyses that are materials-independent. A phase diagram is presented. We further show that the above three magnetic topological states can be indeed implemented in two-dimensional (2D) materials ScLiCl$_5$, LiScZ$_5$ (Z=Cl, Br), and ScLiBr$_5$, respectively. The ferromagnetism in these 2D materials is microscopically revealed from $p$ electrons of halogen atoms. This present study opens a door to explore the exotic topological states as well as quantum magnetism from $p$-orbital electrons by means of the materials-independent approach.  

\bf{Keywords:} $p$-orbital magnetism, square lattice, topological states
\end{abstract}

\pacs{}
\maketitle

\section{Introduction}
In two-dimensional (2D) systems, the coexistence of magnetism and nontrivial topological states can induce many novel physical phenomena. A typical example is the quantum anomalous Hall effect (QAHE), where the combination of ferromagnetism and topological insulator can generate dissipationless edge states at boundaries~\cite{Haldane1988,Onoda2003,Liu2008,He2018,Liu2016,Kou2015}. The quantized Hall conductivity is carried by the edge states, which is robust against disorders and impurities. Owing to the dissipationless chiral edge states, QAHE would have potential applications in ultralow-power consumption spintronic devices~\cite{Wu2014}. Thus, the search for materials with QAHE has attracted extensive interests~\cite{Liu2008,Wu2008,Yu2010,Chang2013,Chang2013a,Si2017}. Since the seminal work of Haldane~\cite{Haldane1988}, the honeycomb lattice is thought to be a proper platform to realize the QAHE, e.g., several ferromagnetic transition metal trihalides with honeycomb lattice were proposed to be candidates for the implementation of QAHE~\cite{He2016,Huang2017,He2017,Sun2018,Wang2018,You2019,Li2020}. Besides, the experimental observation of QAHE was realized in magnetic atom doped systems~\cite{Chang2013,Chang2015,Ou2017,Mogi2015}, and recently in the few layers of magnetic semiconductor MnBi$_2$Te$_4$~\cite{Li2019,Deng2020,Ge2020,Liu2020} with triangular lattice. In these materials, $d$ orbitals of the transition metal play important roles in realizing QAHE. Two interesting questions then arise: whether can the QAHE be realized in other lattices, such as square lattice? Can the QAHE be obtained in materials with $p$ orbitals? The study on these questions cannot only give a further understanding of topological states and quantum magnetism, but also offer new family of materials to search for possible room temperature QAHE.    

In this work, we address these appealing issues by revealing a square lattice with the space group P/4n (No.85) that can accommodate three different $p$-orbital magnetic topological states, i.e. the fully spin-polarized nodal loop semimetal, QAHE and ferromagnetic semiconductor. These three quantum states can be obtained by the symmetry and $k\cdot p$ model analysis, which can be implemented in 2D materials ScLiCl$_5$, LiScZ$_5$ (Z=Cl, Br), and ScLiBr$_5$. It is shown that the ferromagnetism in these 2D materials is attributed to $p$ orbitals. Our findings provide a new mechanism of magnetic topological states from $p$-orbital electrons on square lattices, and also present a novel family of 2D magnetic topological materials with high Chern number. 

\section{Three magnetic topological states on square lattice}
\begin{figure}[tbp]
  \centering
  \includegraphics[scale=0.6,angle=0]{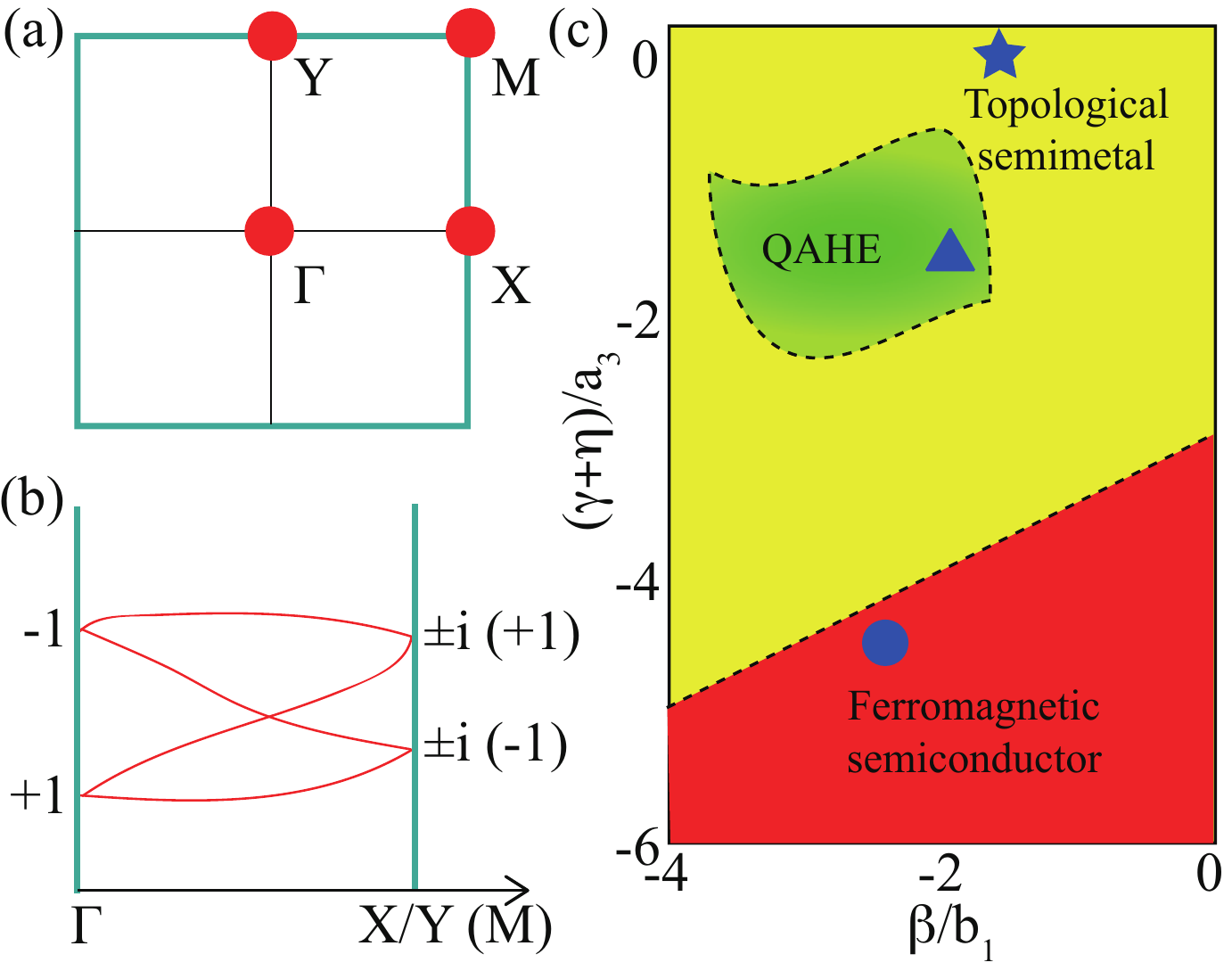}\\
  \caption{The square lattice with three magnetic topological states. (a) Schematic illustration of the double degeneracy at high symmetry points (red dots). (b) Schematic depiction of hourglass dispersion along $\Gamma-X/Y$ and $\Gamma-M$ high symmetry lines. The labels indicate the eigenvalues of $\widetilde{M}_z$. (c) Schematic phase diagram with respect to the parameters $(\gamma+\eta)/a_3$ and $\beta/b_1$ in Eq.~(\ref{kp}), where yellow, red and green regions represent topological semimetal, topologically trivial ferromagnetic semiconductor and QAHE states, respectively.}\label{diagram}
\label{fig1}
\end{figure}

Let us consider a square lattice with the space group of P/4n (No.85) for $p$ orbitals. There are three generators: four-fold rotation symmetry $C_4$: $(x,y,z)\rightarrow(y,-x,z)$, roto-inversion symmetry $\widetilde{I}$: $(x,y,z)\rightarrow(\frac{1}{2}-x,\frac{1}{2}-y,-z)$ and glide mirror symmetry $\widetilde{M}_z$: $(x,y,z)\rightarrow(\frac{1}{2}+x,\frac{1}{2}+y,-z)$. The symmetry protected double degeneracy appears at high-symmetry points in the absence of SOC as shown in Fig.~\ref{fig1}(a). Without SOC, the spin and orbital parts of the electronic wave functions are decoupled, and hence all crystalline symmetries are preserved for each spin channel separately like spinless particles. For the whole BZ, the glide mirror $\widetilde{M}_z$ is preserved. The high symmetry points $\Gamma$, $X$, $Y$ and $M$ are invariant under the combined operation $T\widetilde{M}_z$. We note that $(T\widetilde{M}_z)$$^2$ = $T_{110}$, where $T^2=1$ for the spinless case, and $T_{110}=e^{-ik_x-ik_y}$ represents the translation by one unit cell along the [110] direction. At $X$ point, we have $k_x=\pi$ and $k_y=0$, while at $Y$ point, we have $k_x=0$ and $k_y=\pi$. Consequently, $(T\widetilde{M}_z)$$^2$=-1 for $X$ and $Y$ points. This antiunitary operator thus generates a Kramers-like double degeneracy at $X$ and $Y$ points. One may note that $M$ point is invariant under both $C_4$ and $\widetilde{I}$. The commutation relation between $C_4$ and $\widetilde{I}$ is given by $C_4\widetilde{I}=T_{0\overline{1}0}\widetilde{I}C_4$, where $T_{0\overline{1}0}=e^{ik_y}$. At $M$ point, we have $k_x = \pi$ and $k_y = \pi$; hence, $T_{0\overline{1}0}$ = -1. As a result, for any energy eigenstate $|u\rangle$ with $C_4$ eigenvalue $E_z$, it must have a degenerate partner $\widetilde{I}|u\rangle$ with $C_4$ eigenvalue -$E_z$. This proves that the double degeneracy at $M$ is guaranteed by the symmetry. At $\Gamma$ point $p_x$ and $p_y$ orbitals should be degenerate, while $p_z$ orbitals are not. 

Now we turn to discuss the possible band crossing at high symmetry lines. Consider $\Gamma-X$ and $\Gamma-M$ lines, which are invariant under $\widetilde{M}_z$. The Bloch states along $\Gamma-X$ can be chosen as the eigenstates of $\widetilde{M}_z$ with eigenvalues $E_z=\pm e^{-ik_x/2}$. The glide eigenvalues are $\pm i$ at $X$ and $\pm1$ at $\Gamma$. Because $\Gamma$ and $X$ are both time-reversal-invariant momenta, a Kramers pair has eigenvalues $(+i, -i)$ at $X$, and yet it has (+1, +1) or (-1, -1) at $\Gamma$. This suggests that the pairs must switch partners when going from $X$ to $\Gamma$, and the switching leads to the hourglass-type dispersion, as schematically shown in Fig.~\ref{fig1}(b). A similar analysis applied for path $\Gamma-Y$ shows that the band crossing is also hourglass-type along $\Gamma-Y$. The Bloch states along $\Gamma-M$ have the eigenstates of $\widetilde{M}_z$ with eigenvalues $E_z=\pm e^{-ik_x/2-ik_y/2}$. At $\Gamma$ (0, 0) point, the eigenvalues are (+1, -1), while at $M$ ($\pi$, $\pi$) point the eigenvalues are (-1, +1). Focusing on the middle of two bands, they have opposite eigenvalues, and their ordering is inverted between $\Gamma$ and $M$. As a result, there must be a cross along the $\Gamma-M$ path, and the crossing point is protected by $\widetilde{M}_z$ [Fig.~\ref{fig1}(b)]. Because $\Gamma$ and $M$ both are time-reversal-invariant momenta, a Kramers pair with eigenvalues +1 (or -1) is degenerate, leading to the crossing point of hourglass-type. The crossing point may trace out a nodal loop centered at $\Gamma$.

To characterize the above discussed band crossing, we construct an effective $k\cdot p$ model for the low-energy band structure on square lattice. We first consider the case without SOC. The four states at $\Gamma$ point correspond to two 2D irreducible representations $E_u$ and $E_g$. The model should respect the following symmetries: the fourfold rotation $C_{4z}$, and the mirror symmetry $M_z$. Expanding up to the $k$ quadratic order, we find that the effective Hamiltonian takes the form of
\begin{gather*}
H_{0}=
\begin{pmatrix}
a(k) & -ic(k) &0&0 \\
ic(k)) & a(k)&0&0\\
0&0&b(k)&-id(k)\\
0&0&id(k)&b(k)
\end{pmatrix}
,
\end{gather*}
where $a(k)=a_1+a_2k^2$, $b(k)=b_1+b_2k^2$, $c(k)=a_3+a_4k^2$, $d(k)=b_3+b_4k^2$, $k^2=k_x^2+k_y^2$, and the parameters $a_i$ and $b_i$ (i=1,2,3,4) are real. Considering a fully spin polarized ferromagnet, the inclusion of SOC gives additional contribution to the above Hamiltonian, which can be treated as a perturbation due to the relatively weak SOC strength. We find that the SOC term up to the leading order takes the following form:
\begin{gather*}
H_{\rm SOC}=
\begin{pmatrix}
\alpha & \gamma-\gamma i &0&0 \\
\gamma+\gamma i &\alpha &0&0\\
0&0&\beta&\eta-\eta i\\
0&0&\eta+\eta i&\beta
\end{pmatrix}
,
\end{gather*}
with real parameters $\alpha$, $\beta$, $\gamma$ and $\eta$. Thus, the total Hamiltonian reads
\begin{equation}\label{kp}
H=H_0+H_{\rm SOC}.
\end{equation}

Due to the two-fold degeneracy at $\Gamma$ point in the presence of SOC, we can obtain $a_1=b_1$ and $a_3=b_3$. If we fix $\beta=-\alpha$, a schematic phase diagram of the ratio of $(\gamma+\eta)/a_3$ as a function of $\beta/b_1$ can be drawn as shown in Fig.~\ref{diagram}(c).
Obviously, the SOC term will break the degeneracies at $\Gamma$ point. It may also affect the degeneracy of the nodal loop, i.e. the nodal loop can be preserved with its shape and size changed slightly or the nodal loop vanishes with a band gap opened, and the system becomes QAH insulators or topologically trivial ferromagnetic semiconductors. From Fig.~\ref{diagram}(c), one may note that topologically trivial ferromagnetic semiconductor appears in the region where the SOC parameters $\beta$ and $\gamma+\eta$ has relative large absolute values, and topologically nontrivial states including nodal loop semimetal and topological insulator (QAHE) depend on the relationship and competition between $\beta$ and $\gamma+\eta$. For the QAHE, the band inversion occurs.

\section{Magnetic Topological Materials with Square Lattice}
We now present several 2D material examples to implement the above different topological states, i.e. topological semimetal, topologically trivial ferromagnetic semiconductor and QAHE states as indicated in Fig.~\ref{fig1}(c). These 2D materials with the formula unit XYZ$_5$ possess square lattice with the space group of P/4n, where X atoms occupy the Wyckoff position 2$b$(0; 0; 0.5), Y atoms occupy the Wyckoff position 2$c$(0.5; 0; 0.56383), and Z atoms occupy the Wyckoff positions 8$g$(0.20444, 0.11338, 0.59774) and 2$c$(0.5, 0, 0.40678) as shown in Fig.~\ref{fig2}(a). It is interesting to mention that several bulk materials with similar structures had been synthesized and extensively studied in the last decades~\cite{Kierkegaard1970,Tachez1981,Lezama1989,Amos1998,Carretta2002,Carretta2003,Kiani2016}. 

\subsection{Fully Spin-polarized Nodal Loop Semimetal in ScLiCl$_{5}$}
\begin{figure}[tbp]
  \centering
  \includegraphics[scale=0.42, angle=0]{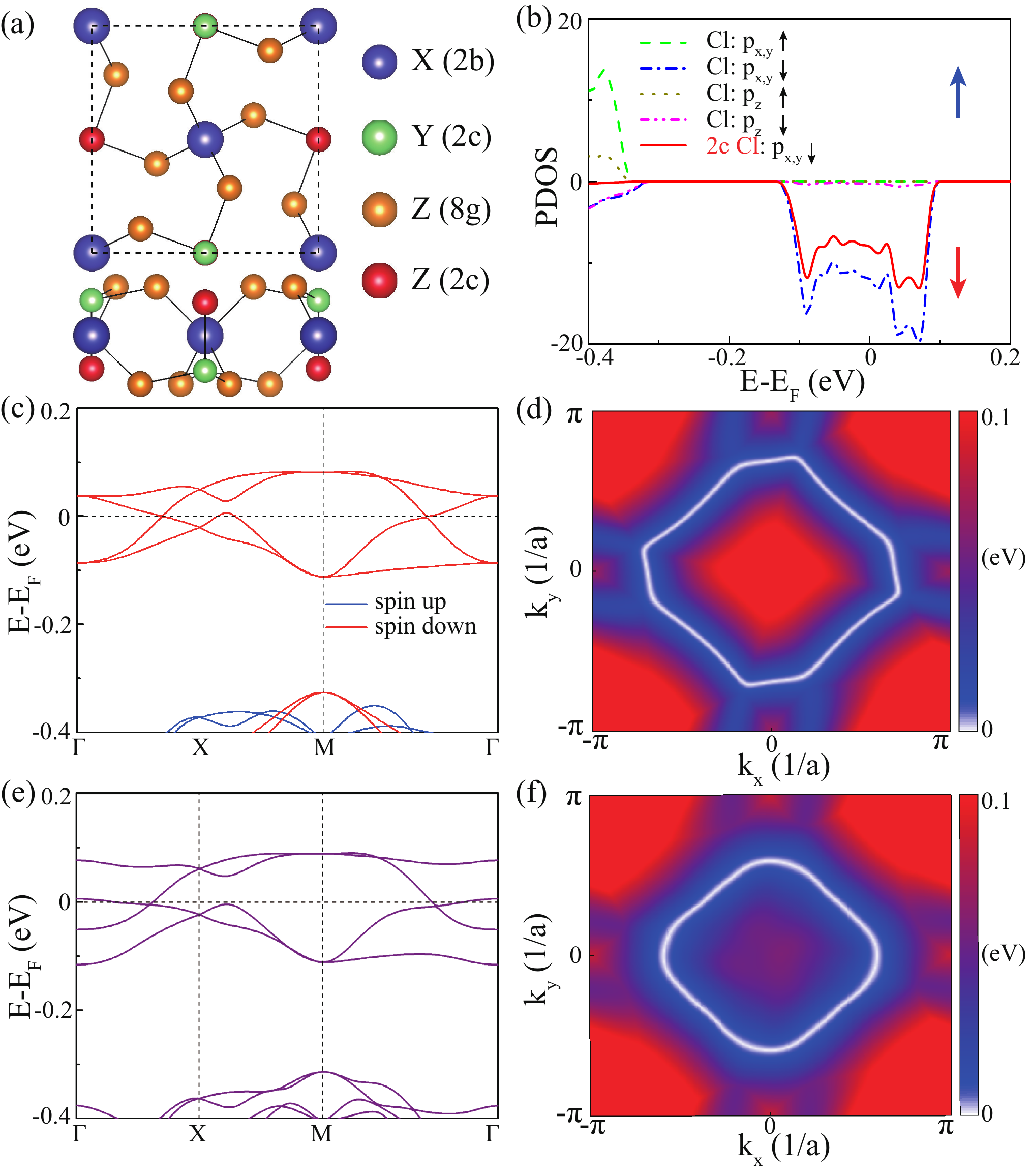}\\
  \caption{Fully spin-polarized nodal loop semimetal ScLiCl$_5$ monolayer. (a) Top and side views of 2D materials XYZ$_5$, where X atoms occupy the Wyckoff position 2$b$(0; 0; 0.5) colored in blue, Y atoms occupy the Wyckoff position 2$c$(0.5; 0; 0.56383) colored in green, and Z atoms occupy the Wyckoff positions 8$g$(0.20444, 0.11338, 0.59774) (orange) and 2$c$(0.5, 0, 0.40678) (red). (b) Partial density of states, (c) band structure and (d) the Weyl loop obtained from DFT calculations in the absence of SOC for ScLiCl$_5$ monolayer. (e) Electronic band structure and (f) the Weyl loop obtained from DFT calculations of ScLiCl$_5$ with SOC.}\label{fig2}
\end{figure}

The structure of ScLiCl$_5$ monolayer with a square lattice is depicted in Fig.~\ref{fig2}(a). Each primitive cell contains two formula units of XYZ$_5$. To confirm the stability of ScLiCl$_5$ monolayer, its phonon spectra have been calculated. There is no imaginary frequency mode in the whole Brillouin zone, indicating that this monolayer is dynamically stable. The structural stability of ScLiCl$_5$ monolayer is also examined in terms of the formation energy. The obtained negative values of formation energy (the energy difference between XYZ$_5$ and X, Y crystals, Z$_2$ molecule) for XYZ$_5$ monolayers are indicative of an exothermic reaction. Moreover, the thermal stability of ScLiCl$_5$ monolayer is tested by using the molecular dynamics simulation by considering a 3$\times$3$\times$1 supercell of ScLiCl$_5$ with 126 atoms. After being heated at 300 K for 6 ps with a time step of 3 fs, no structural changes occur, indicating that this monolayer is also thermodynamically stable. More details can be found in Supplemental Materials (SM).

The optimized lattice constant of ScLiCl$_5$ monolayer is $a_0$ = 7.9372 \AA. By a spin-polarized calculation, we note that the total spin magnetic moment carried by ScLiCl$_5$ is about 1.53 $\mu_B$ per unit cell, which is mainly attributed to the two Cl atoms on Wyckoff position $2c$ (colored by red in Fig.~\ref{fig2}(a)), whose spin and orbital  moments are of about 0.50 and 0.16 $\mu_B$ per atom, respectively. The ferromagnetism is mainly originated from the $p$ orbitals of Cl atoms, whereas the spin magnetic moment on Sc atom is calculated to be zero. It is well interpreted that, for ScLiCl$_5$, because Li and Sc have one and three valence electrons, respectively, Cl atoms possess unpaired electrons and thus should carry a nonzero spin magnetic moment. To determine the magnetic ground state, we compared the total energies between FM, antiferromagnetic (AFM) and non-magnetic (NM) states. The FM state is found to be more stable than AFM and NM states.

The partial density of states (PDOS) and electronic band structure in the absence of SOC are shown in Figs.~\ref{fig2}(b) and (c), respectively. One observes that the material is a half-metal, with only one spin channel (spin down) being metallic and another spin channel (spin up) being insulating. From the projected density of states (PDOS) as displayed in Fig.~\ref{fig2}(b), one may see that the states around the Fermi energy are fully polarized in the spin down channel, while the spin up channel has a large gap. In addition, the low-energy states are dominated by the $p_{x,y}$ orbitals of the Cl atoms on Wyckoff position $2c$. From Fig.~\ref{fig2}(c), we find two features of the band structure. One is the double degeneracy at high symmetry points $\Gamma$, $X$ and $M$, and another is the linear band-crossing points appearing on the paths $\Gamma$-$X$, and $\Gamma$-$M$. These crossing points are not isolated, and form a nodal loop around $\Gamma$ point as shown in Fig.~\ref{fig2}(d).

In the presence of SOC, the magnetic anisotropy should be considered. In order to determine the easy axis of magnetization, we shall pin down the magnetization direction for FM configurations. By comparing the energies of different magnetization directions, we find that the out-of-plane direction is energetically preferred over the in-plane directions and along that they are isotropic. We have also estimated the Curie temperature $T_C$ for the FM state by using the Monte Carlo simulation based on an effective Hamiltonian
\begin{equation}\label{Hamiltonian}
H_{spin}=\sum_{\langle i,j \rangle}J_1S_i^zS_j^z+\sum_{\langle\langle i,j \rangle\rangle}J_2S_i^zS_j^z,
\end{equation}
where the spin vectors are normalized, the superscripts $i$ and $j$ label the $2c$ Cl sites, $\langle i,j\rangle$ and $\langle\langle i,j \rangle\rangle$ indicate nearest-neighboring and next nearest-neighboring sites, respectively, $J_1$ and $J_2$ are the corresponding FM exchange integrals. The values of $J_1$ and $J_2$ extracted from DFT calculations are -4.572 and -0.161 meV, respectively. The calculated Curie temperature for monolayer ScLiCl$_5$ is about 123 K.

The electronic band structure with SOC for ScLiCl$_5$ monolayer is shown in Fig.~\ref{fig2}(e). It is noted that, the SOC only breaks the degeneracy at $\Gamma$ point, but keeps the degeneracies at $X(Y)$ and $M$ points, which are protected by symmetry. The nodal loop is also preserved with SOC in monolayer ScLiCl$_5$ as shown in Fig.~\ref{fig2}(f). Thus, ScLiCl$_5$ monolayer exhibits a fully spin-polarized nodal loop semimetal (we could call it as nodal loop half-semimetal).
By fitting the two bands near Fermi level, we can obtain the parameters for ScLiCl$_5$: $\beta/a_1=-1.4$ and $(\gamma+\eta)/a_3=0$. Thus, ScLiCl$_5$ locates in the region of topological semimetal as marked by a star in Fig.~\ref{fig1}(c).

\begin{figure}[!htbp]
  \centering
  \includegraphics[scale=0.41,angle=0]{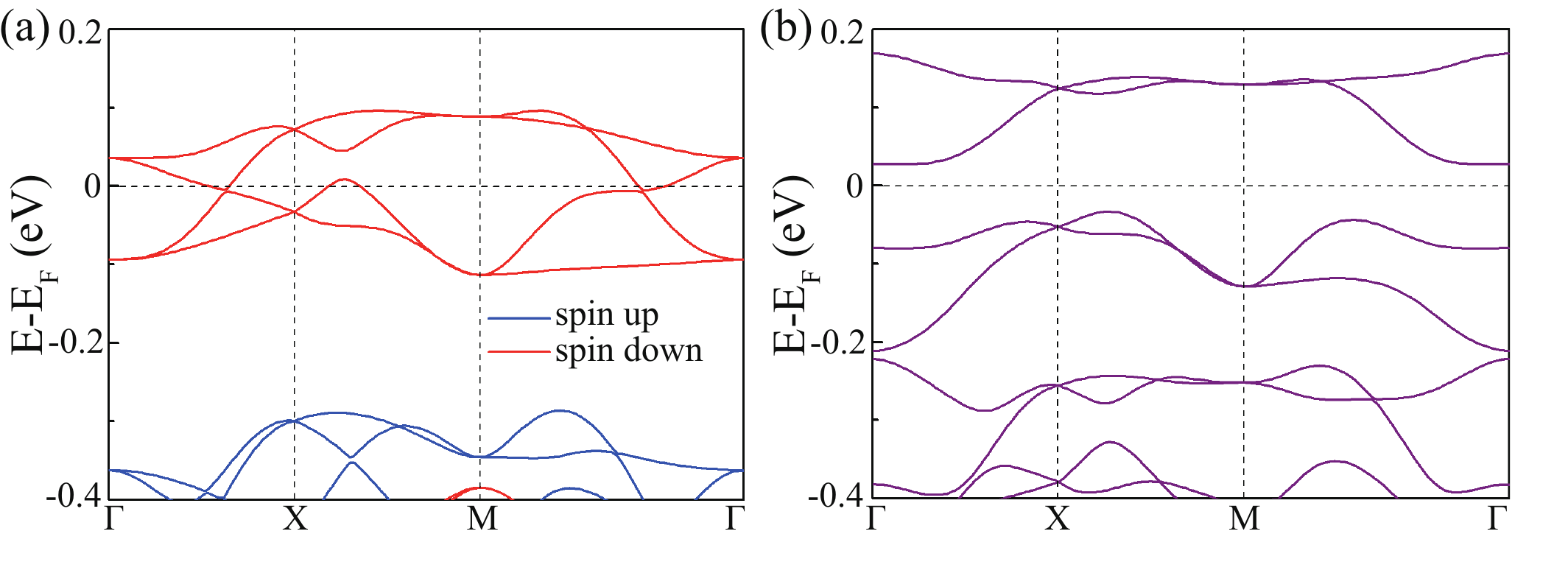}\\
  \caption{Ferromagnetic semiconductor ScLiBr$_5$ monolayer. The electronic band structure (a) without SOC and (b) with SOC.}\label{fig3}
\end{figure}

\subsection{Ferromagnetic Semiconductor in ScLiBr$_{5}$}
The monolayer ScLiBr$_5$ shares the same structure as ScLiCl$_5$, except a larger lattice constant $a=8.4175$\AA. The stability of monolayer ScLiBr$_5$ is also checked by its phonon spectra, molecular dynamics and formation energy, indicating it is feasible in experiment (see SM). The spin-polarized calculation shows that the total spin magnetic moment of ScLiBr$_5$ is about 1.43 $\mu_B$ per unit cell, and the two Br atoms on Wyckoff position $2c$ possess spin and orbital magnetic moments of about 0.46 and 0.34 $\mu_B$ per atom, respectively, whereas spin and orbital magnetic moments on other atoms are negligible. By comparing the total energies between FM, AFM and NM states, the FM state is found to be more stable than antiferromagnetic and non-magnetic states. 
The monolayer ScLiBr$_5$ has the similar band structure in the presence of SOC as shown in Fig.~\ref{fig3}(a). In this case, the monolayer ScLiBr$_5$ possesses a ferromagnetic ground state with out-of-plane magnetization, and the Curie temperature was estimated to be 67 K by the Monte Carlo simulation based on Eq.~(\ref{Hamiltonian}) with $J_1$=-1.990 meV and $J_2$=-0.455 meV. The band gap of about 60.2 meV is opened by SOC for monolayer ScLiBr$_5$, and it turns into a topologically trivial ferromagnetic semiconductor state with a zero Chern number. By fitting the two bands near Fermi level, we can obtain the parameters for ScLiBr$_5$: $\beta/a_1=-2.3$ and $(\gamma+\eta)/a_3=-4.6$, which locates in the region of ferromagnetic semiconductor as marked by a dot in Fig.~\ref{fig1}(c).

\begin{figure}[!htbp]
  \centering
  \includegraphics[scale=0.41,angle=0]{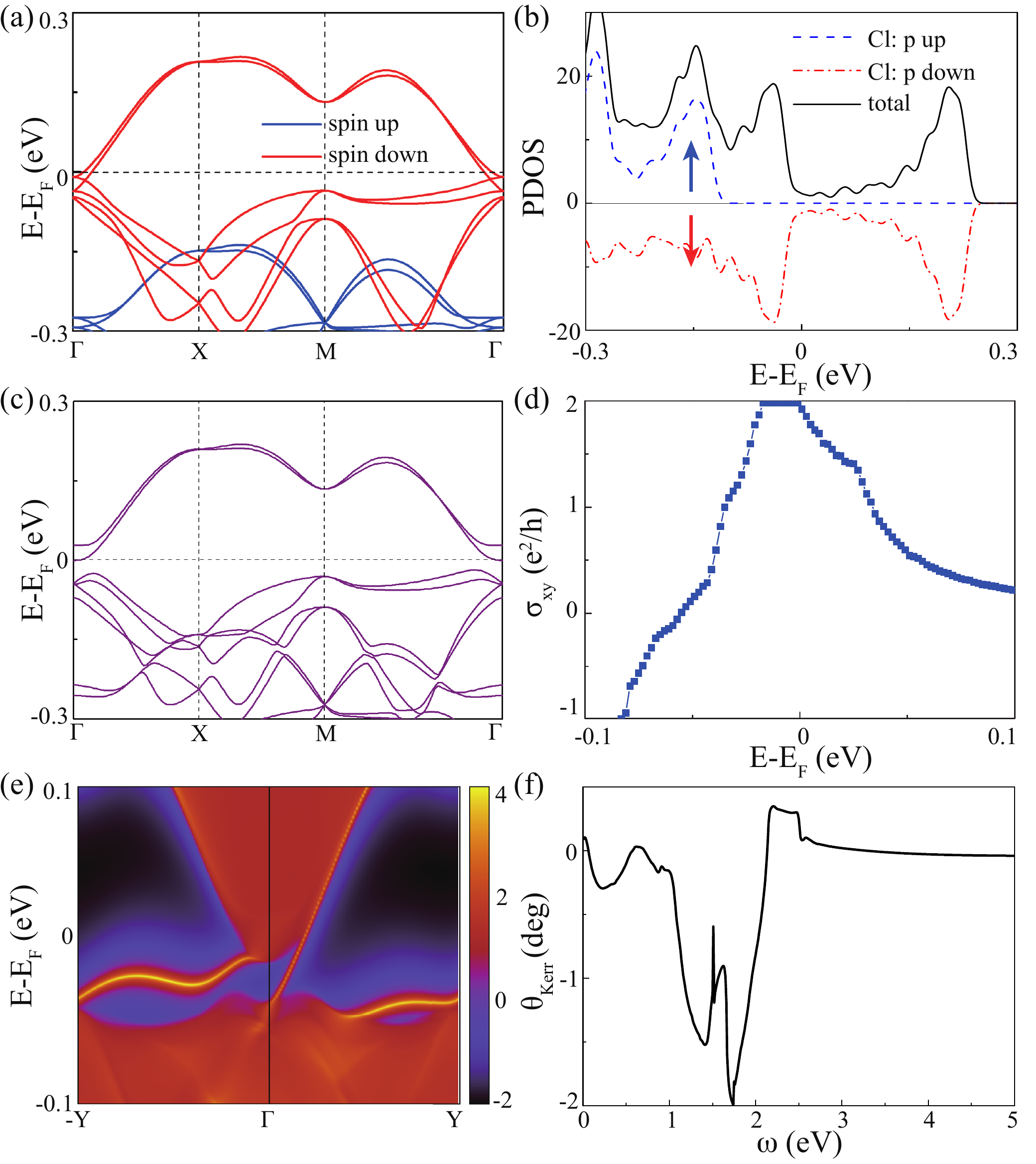}\\
  \caption{QAHE in LiScCl$_5$ monolayer. (a) Band structure in the absence of SOC and (b) partial density of states. (c) The band structure, (d) anomalous Hall conductivity and (e) projected spectrum on (100) surface (line for 2D) with SOC. (f) The Kerr angle $\theta_{Kerr}$ as a function of photon energy $\omega$.}\label{fig4}
\end{figure}

\subsection{QAHE in LiScX$_{5}$ (X=Cl, Br)}
The monolayer LiScX$_5$ (X=Cl, Br) can be obtained by exchanging the positions of Li and Sc atoms in the structures of ScLiCl$_5$ and ScLiBr$_5$ [Fig.~\ref{fig2}(a)]. We shall take LiScCl$_5$ as a prototypal example because LiScBr$_5$ shares very similar features. The optimized lattice constant of LiScCl$_5$ monolayer is $a_0 = 8.0048$ \AA. The stability of monolayer LiScCl$_5$ is also checked by its phonon spectra, molecular dynamics and formation energy (see SM), indicating it is feasible in experiment. By comparing the formation energy, one may find that ScLiCl$_5$ (-1.52 eV) is little more stable than LiScCl$_5$ (-1.51 eV). By a spin-polarized calculation, we found that the spin magnetic moment carried by LiScCl$_5$ is about 1.54 $\mu_B$ per cell shared by all Cl atoms, while the Cl atoms on Wyckoff position $2c$ have orbital moment (0.03 $\mu_B$) much larger than that (0.01 $\mu_B$) for the Cl atoms on Wyckoff position $8g$. Our calculated results show that the FM state is more stable than antiferromagnetic and non-magnetic states.

The electronic band structure and density of states (DOS) for monolayer LiScCl$_5$ in the absence of SOC are shown in Figs.~\ref{fig4}(a) and (b), respectively. It can be seen that the monolayer LiScCl$_5$ holds the similar band structure as ScLiCl$_5$ in the presence of SOC except that the band crossing points are closer to the $\Gamma$ point. From the projected density of states (PDOS) as displayed in Fig.~\ref{fig4}(b), we may observe that the states around the Fermi energy are fully polarized in the spin down channel, while the spin up channel has a large gap. The low-energy states are dominated by the $p$ orbitals of all Cl atoms.

In the presence of SOC, the magnetic anisotropy should be considered. By comparing the energies of different magnetization directions, we uncover that the out-of-plane direction is energetically preferred over the in-plane directions, and the energy of out-of-plane magnetization is 1.10 meV lower than that of in-plane magnetization. The Curie temperature is estimated to be 28 K. After tuning on SOC, a gap of about 24.7 meV is opened as shown in Fig.~\ref{fig4}(c).  The topologically nontrivial band structure of LiScCl$_5$ monolayer is characterized by a nonzero Chern number $C$ = 2 with a quantized charge Hall plateau 2e$^2$/h and two gapless chiral edge states connecting the valence and conduction bands as shown in Figs.~\ref{fig4}(d) and (e), respectively. In addition to the QAHE with high Chern number, the magneto-optical Kerr effect, being a kind of non-contact (non-damaging) optical technique, is a powerful tool for measuring the magnetism in 2D materials~\cite{Gong2017,Chang2013a}. It can be seen that a large Kerr angle $\theta_{\rm Kerr}$ is obtained for LiScCl$_5$ monolayer, particularly for photon energies $\omega$ near 1.7 eV as shown in Fig.~\ref{fig4}(f). The maximal Kerr angle for LiScCl$_5$ monolayer is an order of magnitude larger than that for CrGeTe$_3$ monolayer~\cite{Gong2017}, and about 3 times larger than that for bulk Fe~\cite{1968Magneto}. By fitting the two bands near Fermi level, we can obtain the parameters for LiScCl$_5$: $\beta/a_1=-2.0$ and $(\gamma+\eta)/a_3=-1.9$, which locates in the region of the QAHE as marked by a triangle in Fig.~\ref{fig1}(c). The high-Chern-number QAHE with a large band gap of about 113 meV can be also implemented in monolayer LiScBr$_5$ (see SM). 

\begin{table}[htbp]
  \caption{The differences of matrix element squared between two directions of the magnetization ($|\langle o^{-}|L_z|u^{-}\rangle|^2$-$|\langle o^{-}|L_x|u^{-}\rangle|^2$) and ($|\langle o^{+}|L_z|u^{-}\rangle|^2$-$|\langle o^{+}|L_x|u^{-}\rangle|^2$)  in Eq.~(\ref{eq3}), where $o$ and $u$ are occupied and unoccupied orbitals, + and - are majority and minority spin states, respectively.}
  \label{tab:1}
  \setlength{\tabcolsep}{2.6mm}
  \begin{tabular}{cccccccc}
		\hline
		   \ &$p_{x,+}$  \ &$p_{y,+}$  \ &$p_{z,+}$  \  &$p_{x,-}$  \ &$p_{y,-}$ &$p_{z,-}$ \\
		\hline  \hline
		$p_{x,+}$    &0     &1    &0     &0     &-1   &0     \\
		$p_{y,+}$    &1    &0     &-1     &-1     &0   &1   \\
		$p_{z,+}$    &0     &-1     &0     &0     &1  &0   \\
        $p_{x,-}$  &0     &-1     &0     &0     &1  &0     \\
        $p_{y,-}$  &-1     &0     &1    &1    &0   &-1         \\
        $p_{z,-}$  &0     &1    &0     &0     &-1   &0      \\
		\hline
	\end{tabular}
\end{table}

\section{Magnetic single-ion anisotropy}
We shall take ScLiCl$_5$ as an example to discuss the microscopic mechanism of the out-of-plane magnetization and large magnetic anisotropy.

\begin{table}[htbp]
  \caption{Orbital-resolved magnetic single-ion anisotropic energy $E_{\rm SIA}$ of Cl atoms in the fully spin-polarized nodal loop semimetal ScLiCl$_5$, where the dominated $E_{\rm SIA}$ comes from ($p_x$, $p_y$) [or ($p_y$, $p_x$)] matrix element of Cl atoms on Wyckoff position 2$c$.}
  \label{tab:2}
    \setlength{\tabcolsep}{3mm}
  \begin{tabular}{ccccc}
		\hline
		   & &$(p_x, p_y)$  \ &$(p_x, p_z)$  \ &$(p_y, p_z)$ \\
		\hline  \hline
		\multirow{2}*{{\rm Cl}} &$2c$    &3.08     &0.10    &-0.44     \\
		&$8g$    &-0.05    &0.00     &-0.01     \\
		\hline
	\end{tabular}
\end{table}

According to the second order perturbation theory, the magnetic anisotropy from single-ion anisotropy (SIA) can be described as~\cite{Wang1993,Yang2017} 
\begin{equation}
E_{\rm SIA}=\lambda^2\sum_{o,u}\frac{|\langle\psi_u|L_z|\psi_o\rangle|^2-|\langle\psi_u|L_x|\psi_o\rangle|^2}{\epsilon_u-\epsilon_o},
\label{eq3}
\end{equation}
where $\lambda$ is the SOC constant, $L_{z/x}$ represent the angular momentum operators, and $\epsilon_u$ and $\epsilon_o$ are the unoccupied
and occupied energy, respectively. A positive value of $E_{\rm SIA}$ indicates the out-of-plane magnetization, otherwise the in-plane. Equation~(\ref{eq3}) means that the orbitals near the Fermi energy mainly contribute to MAE. By calculating the differences of matrix element squared between two directions of the magnetization for $p$ orbitals according to Eq.~(\ref{eq3}) as shown in Table~\ref{tab:1}, we note that the contributions to MAE from the same spins and from the opposite spins between occupied ($|\psi_o\rangle$) and unoccupied ($|\psi_u\rangle$) states have opposite signs. Positive and negative matrix elements prefer the out-of-plane and in-plane magnetization, respectively. In our systems, the states near the Fermi energy are mainly contributed by the same spin (spin down) of $p_{x,y}$ orbitals, which should prefer an out-of-plane magnetization. 

To confirm the above observation, the orbital-resolved $E_{\rm SIA}$ was calculated for ScLiCl$_5$ monolayer as listed in Table~\ref{tab:2}. It is seen that Li and Sc atoms as well as Cl atoms on Wyckoff position 8$g$ make no contribution to $E_{\rm SIA}$, while the main contribution comes from Cl atoms on Wyckoff position 2$c$ as revealed in Table~\ref{tab:2}. The value of ($p_x$, $p_y$) matrix element is dominated and positive, indicating an out-of-plane magnetization, which is consistent with the above analysis. For monolayers ScLiBr$_5$ and LiScX$_5$ (X=Cl, Br), the same analysis applies and we find that they all prefer the out-of-plane magnetization, which is consistent with our DFT results.

\begin{table}[htbp]
  \caption{The spin $\langle S \rangle$ and orbital $\langle O \rangle$ moments (in $\mu_B$) of Z (or Cr) atoms, magnetic anisotropy (MAE) (in meV) per formula unit between the out-of-plane and in-plane FM configurations, and Curie temperature T$_{c}$ (in K) for XYZ$_5$ compounds as well as CrGeTe$_3$\cite{You2020} for comparison.}
  \label{tab:3}
  \setlength{\tabcolsep}{2.6mm}
  \begin{tabular}{ccccccc}
		\hline
		   \multirow{2}*{Monolayer }  &Wyckoff &\multirow{2}*{$\langle S \rangle$}  \ &\multirow{2}*{$\langle O \rangle$}  \ &\multirow{2}*{MAE}  &\multirow{2}*{T$_c$} \\
		   &position &&&&\\
		\hline  \hline
		 \multirow{2}*{ScLiCl$_5$}     &$2c$    &0.497     &0.159    &\multirow{2}*{3.512}    &\multirow{2}*{123}     \\
		                         &$8g$    &0.066    &0.002     & &     \\
		  \hline
		   \multirow{2}*{ScLiBr$_5$}     &$2c$    &0.460     &0.337    &\multirow{2}*{7.919}  &\multirow{2}*{67}    \\
		                         &$8g$    &0.063    &0.002     & &     \\
		  \hline
                 \multirow{2}*{LiScCl$_5$}     &$2c$    &0.136     &0.028    &\multirow{2}*{0.504} &\multirow{2}*{28}     \\
		                         &$8g$    &0.162    &0.006     & &   \\
		  \hline
                 \multirow{2}*{LiScBr$_5$}     &$2c$    &0.166     &0.069    &\multirow{2}*{5.418}   &\multirow{2}*{20}  \\
		                         &$8g$    &0.142    &0.020     & &     \\
		  \hline
                  CrGeTe$_3$  &$--$ &3.614  &0.004 &1.850  &19\\
		\hline
	\end{tabular}
\end{table}

The atomic SOC is calculated by H$_{SOC} = \lambda \mathbf{S}\cdot\mathbf{L}$, where $\lambda$ related to the atomic number is the coefficient of SOC and $\mathbf{S}$ and $\mathbf{L}$ represent the spin and orbital angular momentum operators, respectively. Although $\lambda$ is small for Cl atom ($\sim$ 331 cm$^{-1}$)\cite{Martin1971}, its orbital moment is large ($\sim$ 0.16 $\mu_B$) as shown in Table~\ref{tab:3}, while in many $d$-orbital magnetic materials though their $\lambda$ is much bigger than that of Cl atom, their orbital magnetic moments are quenched, and thus in our compounds, the atomic SOC of $p$ orbitals is large, and opens a gap of tens meV, which is comparable with and even larger than many $d$-orbital magnetic materials. For example, for monolayer CrGeTe$_3$, its $\lambda$ is 740 cm$^{-1}$\cite{Martin1971}, while its orbital moment is quenched ($\sim$ 0.004 $\mu_B$)\cite{You2020}, leading to a small SOC. Due to the large SOC in our compounds, the obtained single-ion anisotropy is large.

From Table~\ref{tab:3}, it is noted that the estimated Curie temperature for our compounds is higher than that of CrGeTe$_3$, especially for ScLiCl$_5$ and ScLiBr$_5$. Moreover, for ScLiCl$_5$, ScLiBr$_5$ and LiScBr$_5$, their magnetic anisotropies are much higher than CrGeTe$_3$ because the former has larger orbital magnetic momenta. Thus, the $p$-orbital magnetism in our compounds is reliable and stable.

\section{Conclusion}
In this work, we propose the $p$-orbital topological magnetic states on a square lattice with space group P/4n by means of the symmetries and $k\cdot p$ model analyses that are materials-independent. Three currently interested topological states, including topological semimetal, QAHE and topologically trivial ferromagnetic semiconductor, can be obtained on the square lattice, depending on the interplay between different SOC parameters. A phase diagram is presented. As examples, we show that the above three different topological states can be indeed implemented in 2D materials ScLiCl$_5$, ScLiBr$_5$ and LiScCl$_5$ (or LiScBr$_5$), respectively. Furthermore, the ferromagnetism of these 2D ferromagnets is unveiled from the $p$-orbitals of halogen elements, and the microscopic origin of ferromagnetism from $p$ electrons is elaborated. This present study opens a door to explore not only exotic topological states (e.g. nodal loop half-semimetal) but also the quantum magnetism from $p$-orbital electrons in terms of the model and materials-independent analyses.

\section*{Acknowledgements}
This work is supported in part by the National Key R\&D Program of China (Grant No. 2018YFA0305800), the Strategic Priority Research Program of the Chinese Academy of Sciences (Grant No. XDB28000000), the National Natural Science Foundation of China (Grant No.11834014), and Beijing Municipal Science and Technology Commission (Grant No. Z191100007219013). B.G. is also supported by the National Natural Science Foundation of China (Grants No. Y81Z01A1A9 and No. 12074378), the Chinese Academy of Sciences (Grants No. Y929013EA2 and No. E0EG4301X2), the University of Chinese Academy of Sciences (Grant No. 110200M208), the Strategic Priority Research Program of Chinese Academy of Sciences (Grant No. XDB33000000), and the Beijing Natural Science Foundation (Grant No. Z190011).

\section*{AUTHOR CONTRIBUTIONS}
J. Y. You, B. Gu and G. Su designed research. J. Y. You did the calculations and wrote the paper. All authors discussed the results and revised the paper.
\section*{Conflict of interest statement}
None declared.


%

\end{document}